 \documentclass{jpsj2}

\def\runauthor{Takashi \textsc{Shirahata} and Tota \textsc{Nakamura}}

\title{%
A Simultaneous Magneto-Dielectric Phase Transition in RbCoBr$_3$
}

\author{%
Takashi \textsc{Shirahata}
\thanks{Present address:
  Hitachi Medical Corporation,
  2-1, Shintoyofuta, Kashiwa, Chiba, 277-0804 Japan
}
and Tota \textsc{Nakamura}
}

\inst{%
Department of Applied Physics, Tohoku University,
         Aoba, Sendai, Miyagi  980-8579
}

\recdate{\today}

\abst{%
We have modeled magneto-dielectric phase transitions in ABX$_3$-type
layered triangular lattice compounds.
The model consists of a spin system (magnetic transition)
and a lattice system (dielectric transition).
Magnetic exchange interactions are supposed to change with a relative position
between two spins.
This assumption produces an effective coupling between the lattice and the spin.
Applying the nonequilibrium relaxation method,
we have found that a simultaneous magneto-dielectric phase transition occurs 
when energy scales of the spin part and the lattice part coincide.
An intermediate phase like the partial disordered phase disappears.
Both systems relax frustration of the other system cooperatively
and realize ordered states.
}

\kword{%
Magneto-dielectric transition, layered triangular lattice,
nonequilibrium relaxation method, 
relaxation of frustration by lattice distortion
}

\begin{document}
\maketitle

\section{Introduction}

The ABX$_3$-type layered triangular lattice antiferromagnets have been 
attracting great interests both experimentally and theoretically.
\cite{achiwa69,ex62,ex53,shiba,ex55,ex64,ex63,exCnoano,kurata,koseki,%
ex50,morishita51,morishita52}
As the typical compounds we may list CsNiCl$_3$, CsCoCl$_3$, RbMnBr$_3$, 
KNiCl$_3$ and RbCoBr$_3$.
The crystal structure is hexagonal close packed.
Face-sharing BX$_6$ octahedra run along the $c$-axis forming a BX$_3$ chain.
Magnetic B$^{2+}$ ions form an equilateral triangular lattice on the
$c$-plane.  It causes frustration in the exchange interactions.
Exchange interactions along the BX$_3$ chains are much stronger than those on
the plane.  This spin system can be considered as
a quasi-one-dimensional system with frustration on the $c$-plane.

Successive magnetic phase transitions occur in the Ising antiferromagnet on
the two-dimensional triangular lattice with ferromagnetic next-nearest-neighbor
interactions. \cite{ex62,ex55,ex64}.
A low-temperature magnetic structure is the ferrimagnetic state
(we abbreviate it with ``Ferri'' hereafter).
There exists a partially-disordered (PD) phase between the paramagnetic (Para)
phase and the ferrimagnetic phase.
One of three sublattices is completely disordered in this phase.
The other sublattices take antiferromagnetic configurations.
The PD phase is also considered to exist 
in the layered three-dimensional model.\cite{shiba,ex55,ex63,kurata,koseki}
Since no anomaly of the specific heat was observed in the experiment
\cite{exCnoano} 
between the ferrimagnetic phase and the PD phase,
there is a possibility that this change of the magnetic structure is a
crossover between the PD-like state and the ferrimagnetic-like state.
\cite{koseki}

\begin{figure}[t]
\begin{center}
\includegraphics[width=2.6cm]{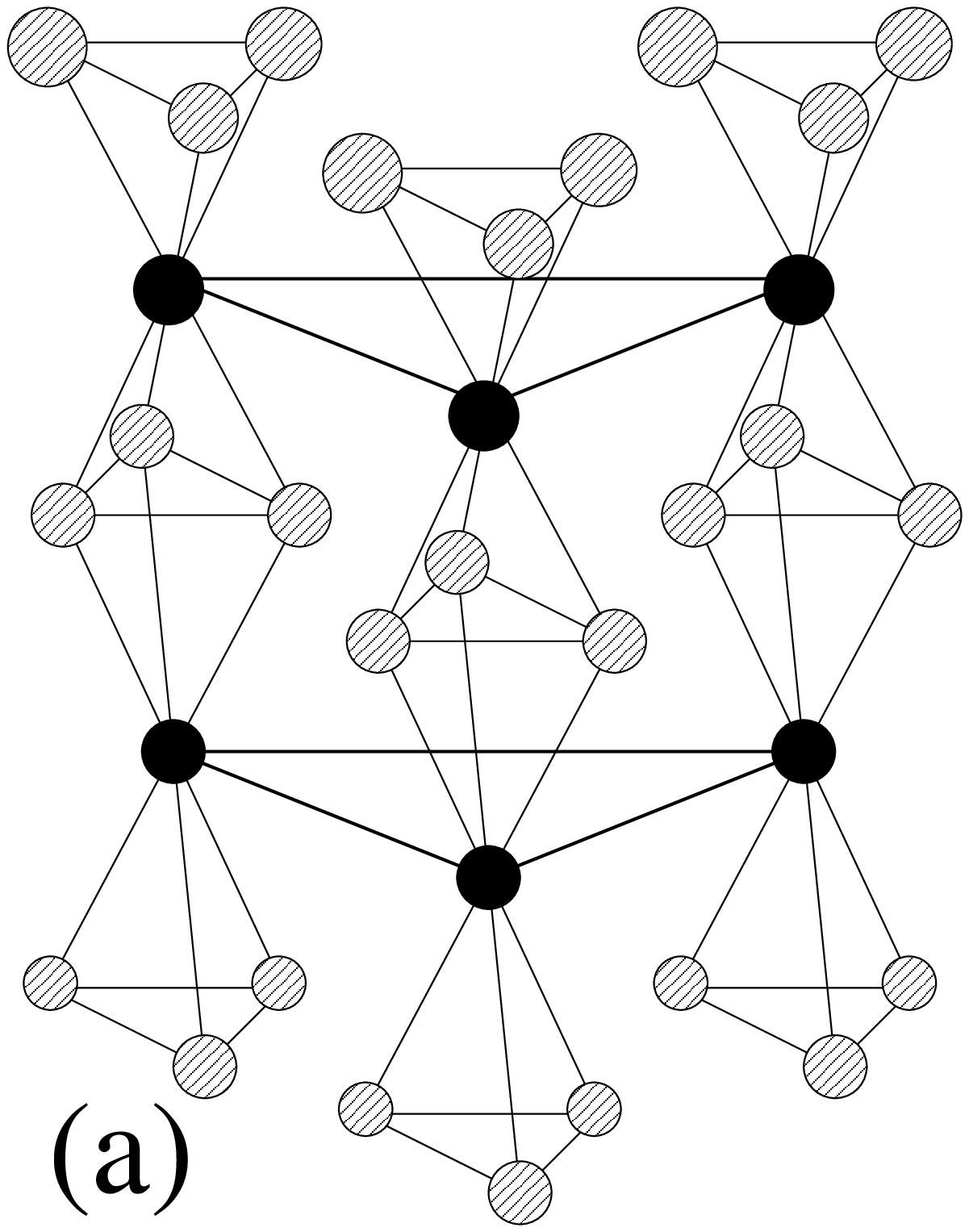}
\includegraphics[width=2.6cm]{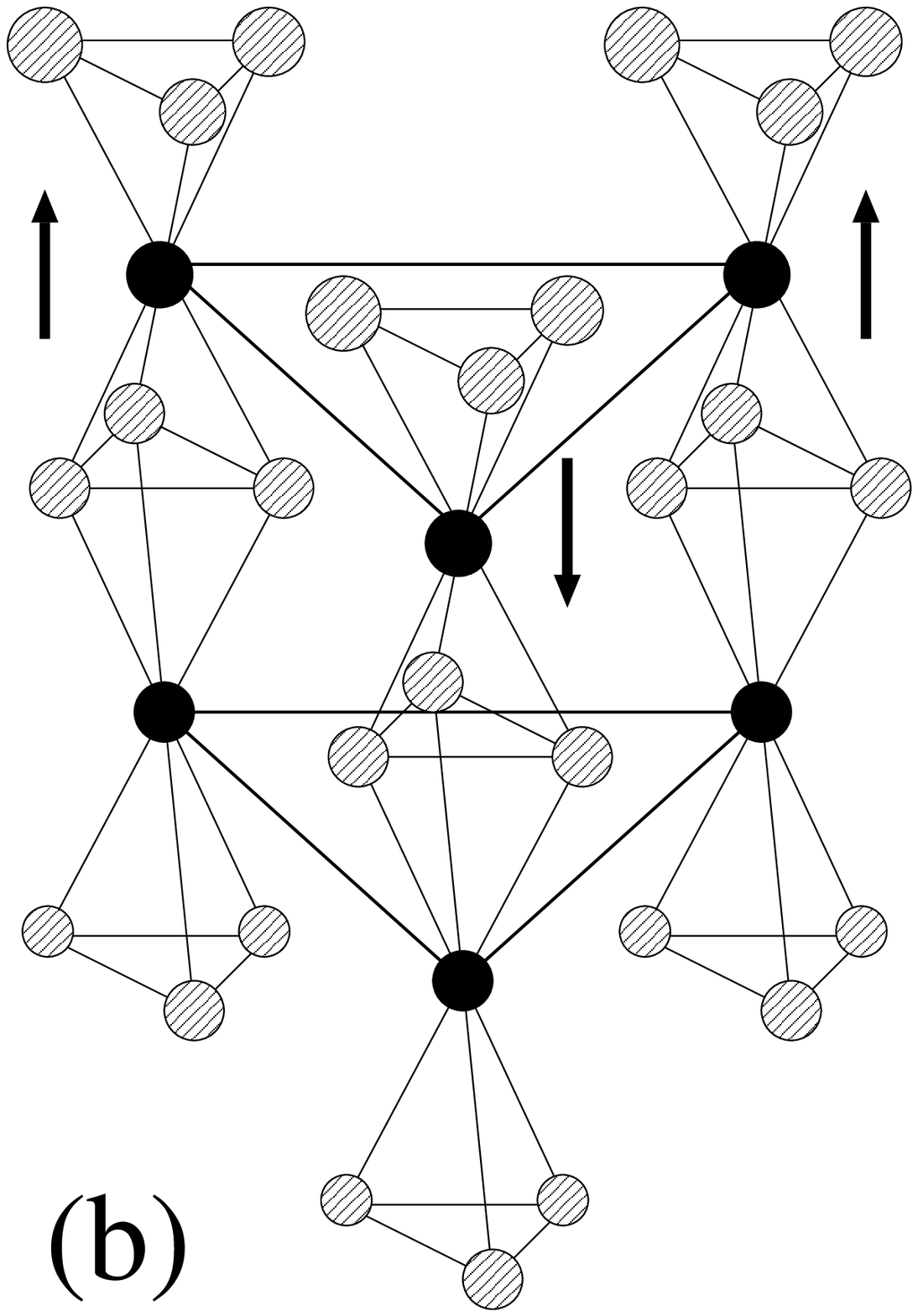}
\includegraphics[width=2.6cm]{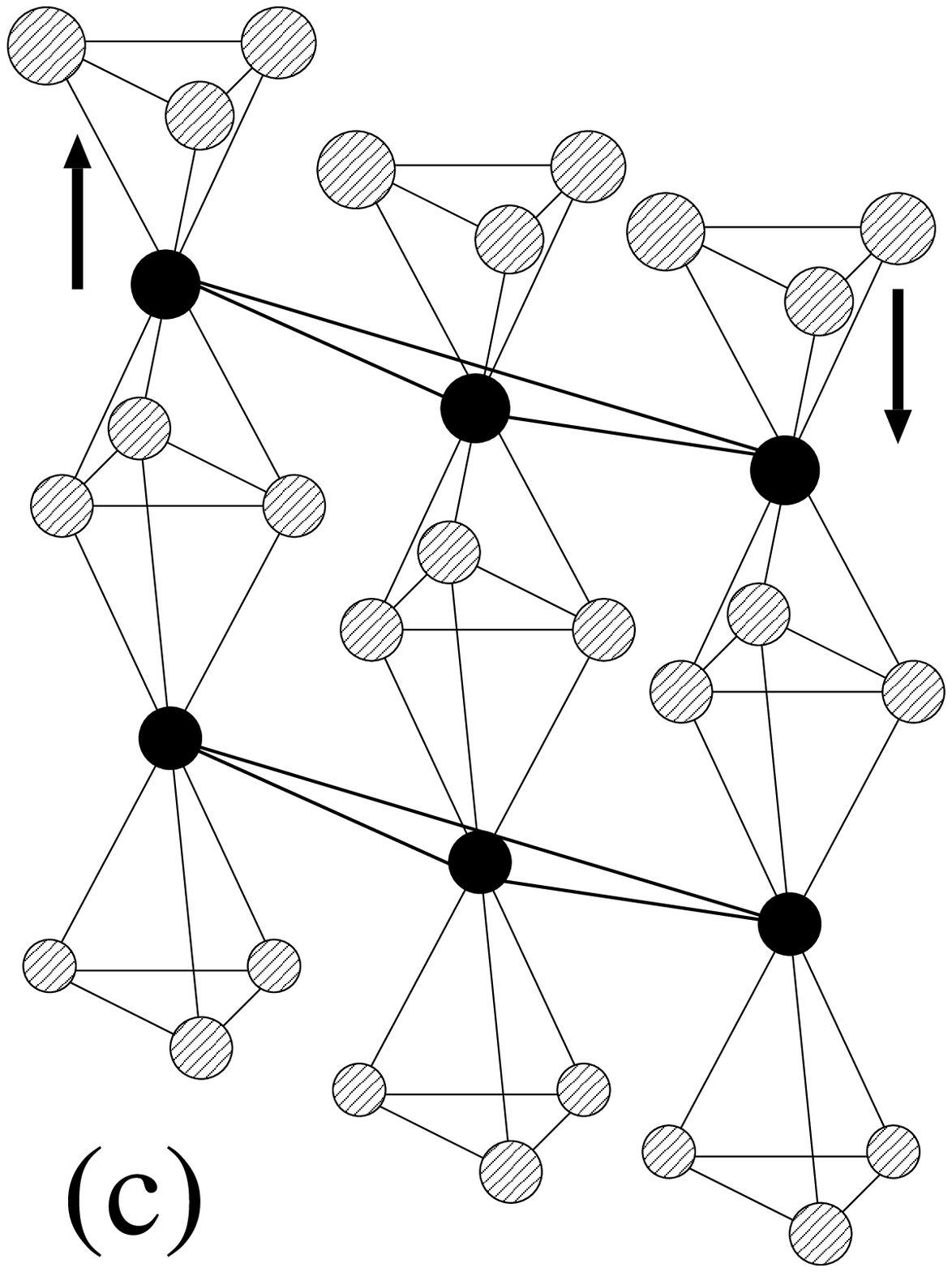}
\end{center}
\caption{Typical crystal structures in ABX$_3$-type compounds.
Solid circles depict magnetic B$^{2+}$ ions and open circles depict X$^-$ ions.
(a) A symmetric structure at high temperatures. The space group is $P6_c/mmc$.
We call this structure ``lattice Para'' in this paper.
(b) A room-temperature KNiCl$_3$ structure. The space group is $P6_3cm$.
We call this structure  ``lattice  Ferri''.
(c) Another low temperature structure. The space group is $P\bar{3}c1$.
We call this structure ``lattice PD''.}
\label{fig:lattice}
\end{figure}

A typical lattice structure of the ABX$_3$-type compounds
at high temperatures is shown in Fig. \ref{fig:lattice} (a).
The space group is $P6_3/mmc$.
The structure is sometimes referred to as a CsNiCl$_3$-type.
Magnetic ions forming an equilateral triangular lattice
sit on a level plane ($c$-plane).
We call this structure ``lattice Para'' in this paper because it is a
symmetric structure at high temperatures like the paramagnetic state.
When the temperature is lowered, structural phase transitions occur.
Each BX$_3$ chain shifts upward or downward keeping the relative distance 
between atoms in a chain unchanged.
Some compounds take a structure in which two of three sublattices on the
triangular lattice shift upward while the other one shifts downward as
shown in Fig. \ref{fig:lattice} (b).
The space group is $P6_3cm$.
It is a room temperature structure of KNiCl$_3$.
Since the displacement is $\uparrow$-$\uparrow$-$\downarrow$ type, we refer to
this structure as ``lattice Ferri'' in this paper.
Other compounds take a structure with one sublattice shifting upward,
one shifting downward and the rest unchanged as shown in 
Fig. \ref{fig:lattice} (c).
The space  group is $P\bar{3}c1$.
We refer to this structure as ``lattice PD'' since the displacement is 
$\uparrow$-$\downarrow$-0.

It is known that each BX$_3$ chain possesses negative charge.
When a structural phase transition occurs,
it can be observed by dielectric polarization.
For example, a structure of the room-temperature KNiCl$_3$ family
(Fig. \ref{fig:lattice} (b)) induces macroscopic polarization. 
This is a distinct difference compared with the other structures in 
Fig. \ref{fig:lattice}.
Morishita et al. \cite{morishita51} classified structural phase
transitions in ABX$_3$-type compounds by the polarization observation.
A structural phase transition and a magnetic phase transition occur
at different temperatures.
A structural phase transition usually occurs at a higher temperature.
This is because an energy scale of the lattice system is
larger than that of the spin system.
A lattice changes its structure from the symmetric one to a distorted one.
A magnetic state remains paramagnetic.
When the temperature is lowered, a magnetic phase transition occurs from
the paramagnetic state to the PD state or to the ferrimagnetic state.

Recently, Morishita et al. found that a structural phase transition and a
magnetic phase transition occur at the same temperature in RbCoBr$_3$.
\cite{morishita52}
The magnetic susceptibility shows a peak at 37K, which suggests that a
magnetically ordered state is realized at lower temperatures.
The magnetic state has not been identified.
The structural phase transition is observed at the same temperature by 
a peak of the dielectric constant.
The low temperature phase exhibits finite spontaneous polarization observed
by a pyroelectricity measurement and a $D$-$E$ hysteresis measurement.
A possible lattice structure with finite polarization is the
room-temperature KNiCl$_3$ structure shown in Fig. \ref{fig:lattice} (b). 
Therefore, it is conjectured that the lattice changes its structure from the 
CsNiCl$_3$-type to the room-temperature KNiCl$_3$-type at this temperature.

In this paper, we make it clear why both magnetic and structural
phase transitions occur simultaneously.
We propose a model Hamiltonian which explains this phenomenon 
in \S \ref{sec:model}.
The model consists of a magnetic part and a lattice part with an effective
coupling between them.
We perform nonequilibrium relaxation (NER) analyses on this Hamiltonian.
A brief review on this method is given in \S \ref{sec:method}.
Numerical results are presented in \S \ref{sec:results}.
Summary and discussions are given in \S \ref{sec:discussion}.
We obtain a phase diagram with regard to a temperature versus a ratio of
energy scales between a magnetic part and a lattice part.
A possible scenario which explains the simultaneous magneto-dielectric
transition is given.

\section{Model}
\label{sec:model}

We construct a model Hamiltonian in order to explain the magneto-dielectric 
phase transition in RbCoBr$_3$.
First, we define state variables.
One lattice point possesses a spin variable $S_{ij}$ and a lattice 
variable $\sigma_{ij}$.
Here, the subscript $i$ denotes a site in the $c$-axis, while $j$ denotes
a site on the $c$-plane.
Since the magnetic ion Co has a strong uniaxial anisotropy,
we treat it with an Ising spin as $S_{ij}=\pm 1$.
For the lattice system, we define $\sigma_{ij}$ by a displacement
from the symmetric lattice point realized in the high-temperature phase.
It is known experimentally that Co ions move only along the $c$-axis.
Therefore, we suppose for simplicity that $\sigma_{ij}$ 
takes $+1$, $-1$ or 0 depending on whether an ion shifts upward, downward or
remains at the symmetric point.
There are six different states at each lattice point.

We introduce the following Hamiltonian consisting of a lattice Hamiltonian
and a spin Hamiltonian.
\begin{equation}
\mathcal{H}=\mathcal{H}_\mathrm{lattice} + \mathcal{H}_\mathrm{spin}
\end  {equation}
A lattice Hamiltonian is supposed to take an expression of elastic energy
with regard to the lattice variables.
A spring constant is denoted by $J^\mathrm{L}_{(c,1,2)}$.
\begin{eqnarray}
\mathcal{H}_\mathrm{lattice}&=&
 \sum_{i,j}
J_c^\mathrm{L} (\sigma_{ij}-\sigma_{(i+1)j})^2 
\nonumber \\
&+&
\sum_i 
\sum_{\langle jk \rangle}^\mathrm{n.n.} 
J_1^\mathrm{L} (\sigma_{ij}-\sigma_{ik})^2
\nonumber \\
&+&
\sum_i 
\sum_{ \langle jl \rangle}^\mathrm{n.n.n.} 
J_2^\mathrm{L} (\sigma_{ij}-\sigma_{il})^2
\end  {eqnarray}
Here, $J_c^\mathrm{L}$, $J_1^\mathrm{L}$ and $J_2^\mathrm{L}$
are spring constants of the nearest-neighbor pairs in the $c$-axis,
the nearest-neighbor pairs on the $c$-plane and
the next-nearest-neighbor pairs on the $c$-plane, respectively. 
$\langle jk\rangle$ denotes a pair in the nearest neighbor and
$\langle jl\rangle$ denotes a pair in the next-nearest neighbor.
It is known experimentally \cite{katoprivate} that lattice dimerization 
within a CoBr$_3$-chain is negligible compared with the whole chain shift.
The lattice is considered to be very hard only along the $c$-axis.
This evidence can be modeled by setting $J_c^\mathrm{L}$ take a positive
value and larger than the other spring constants.
The lattice system possesses a quasi-one-dimensional aspect.
We have also set $J_1^\mathrm{L} < 0$ and $J_2^\mathrm{L} > 0$
in order to realize an $\uparrow$-$\uparrow$-$\downarrow$ conformation
(lattice Ferri) at low temperatures.
This choice is consistent with an exclusion volume effect.
The following parameters are used in our simulations:
\begin{equation}
J_c^\mathrm{L}=5,
J_1^\mathrm{L} = -1,
J_2^\mathrm{L} = 0.1.
\end  {equation}
The temperature is scaled by $|J_1^\mathrm{L}|$.

A spin Hamiltonian is defined by the following expression with
the exchange integrals dependent on the lattice variables.
\begin{eqnarray}
\mathcal{H}_\mathrm{spin}=
&-&\sum_{i,j} 
(J_c^\mathrm{S}-\Delta J_c^\mathrm{S}|\sigma_{ij}-\sigma_{(i+1)j}|) 
S_{ij}S_{(i+1)j}
\nonumber \\
&-&
 \sum_i 
\sum_{\langle jk \rangle}^\mathrm{n.n.} 
(J_1^\mathrm{S} - \Delta J_1^\mathrm{S}|\sigma_{ij}-\sigma_{ik}|)
                                 S_{ij}S_{ik}
\nonumber \\
&-&
\sum_i 
\sum_{\langle jl \rangle}^\mathrm{n.n.n.} 
(J_2^\mathrm{S} - \Delta J_2^\mathrm{S}|\sigma_{ij}-\sigma_{il}|)
                                 S_{ij}S_{il}
\end  {eqnarray}
$J_c^\mathrm{S}$, $J_1^\mathrm{S}$ and $J_2^\mathrm{S}$ are
exchange integrals in the high-temperature symmetric lattice structure.
We have supposed that deformation of the lattice always decreases
magnitudes of exchange integrals. 
Since spins interact via X ions by the super-exchange mechanism, overlap of
electron orbitals determines strength of the interaction.
The overlap decreases when an relative angle along the exchange path
has changed even though the direct distance becomes short. 
Therefore, it is natural to consider that magnitudes of the 
exchange integrals decrease when the lattice is deformed.
For simplicity,
the decrease is supposed to be proportional to an absolute value of the
difference of lattice variables: $|\sigma_{ij}-\sigma_{i'j'}|$, and a
coefficient is $\Delta J^\mathrm{S}_{(c, 1, 2)}$.
The nearest-neighbor interactions within the $c$-plane are supposed to be 
antiferromagnetic and the next-nearest-neighbor interactions are supposed
to be ferromagnetic in order to realize the ferrimagnetic state in the
ground state.
The interactions along the $c$-axis are set ferromagnetic and stronger than
those within the $c$-plane.
We define a ratio of $J^\mathrm{S}_1$ to $J^\mathrm{L}_1$
by $\kappa$ and use the following parameters in our simulations:
\begin{eqnarray}
&&
J_c^\mathrm{S}= 5\kappa,
J_1^\mathrm{S} =-\kappa,
J_2^\mathrm{S} = 0.1\kappa,
~~(\kappa\equiv J_1^\mathrm{S}/J_1^\mathrm{L})
\\
&&
\Delta J_{(c,1,2)}^\mathrm{S}=0.2 J_{(c,1,2)}^\mathrm{S}.
\end  {eqnarray}

Experimental estimates for the exchange integrals are 
$J_c^\mathrm{S}\simeq 62$K, 
$J_1^\mathrm{S}\simeq 2.5$K, and 
$J_2^\mathrm{S}\simeq 1$K.\cite{morishitaC}
A value of $J_c^\mathrm{S}$ is estimated by a position of a broad maximum
peak of $\chi_{\parallel}$.
$J_1^\mathrm{S}$ and $J_2^\mathrm{S}$ are estimated using a molecular field
approximation on the layered equilateral triangular lattice antiferromagnet,
\cite{shiba} 
where the successive transitions of Para$\to$ PD $\to$ Ferri occur.
An application of the theory is not trivial because
a direct transition of Para $\to$ Ferri occurs in the present system.
The obtained results may have differences from true values.
Therefore, we consider that our choices of the exchange integrals
are not inconsistent with the experimental situation
except for $J_c^\mathrm{S}$ being relatively small.
Monte Carlo simulations generally become difficult
when a value of $J_c^\mathrm{S}$ is large.
A lattice length along the $c$-axis should be enlarged in accordance with
a value of $J_c^\mathrm{S}$.
It requires much longer CPU time to perform Monte Carlo simulations.
We have chosen a moderate value of $J_c^\mathrm{S}=5\kappa$ in order to
avoid such time-consuming situations.
It is also noted that a value of $\Delta J_{(c,1,2)}^\mathrm{S}$ is very large:
exchange integrals may be 20\%-40\% decreased by a lattice distortion.
This parameter controls an effective coupling between the spin and the lattice.
When it is large, a lattice distortion strongly influences a spin structure
through changes in the exchange integrals.
The spin and the lattice is independent when $\Delta J_{(c,1,2)}^\mathrm{S}=0$.
We have set the parameter large so that we can easily observe the 
spin-lattice cooperative phenomenon.
We consider that an essential part of the present system can be drawn out
by these choices of parameters.

\begin{figure}[t]
\begin{center}
\includegraphics[width=8.0cm]{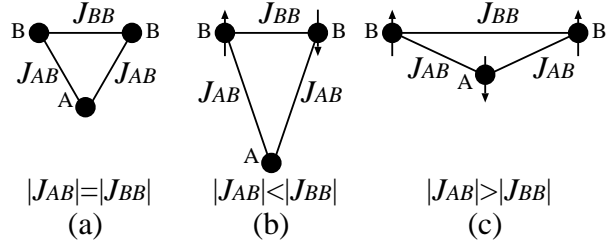}
\end{center}
\caption{(a) A fully-frustrated triangular lattice.
(b) A distorted triangular lattice realized in the room-temperature
KNiCl$_3$-family of Fig. \ref{fig:lattice} (b).
Since the interaction between A and B is weaker than that between B and B,
two B-spins form antiferromagnetic state, while an A-spin takes an up or down
state randomly.  This is the PD configuration.
(c) Another distorted lattice realized in Fig. \ref{fig:lattice} (c).
Since $|J_{AB}|> |J_{BB}|$, A- and B- spins form antiferromagnetic state.
This is the ferrimagnetic configuration.
}
\label{fig:frustrate}
\end{figure}

A parameter $\kappa$ is  controllable and we have carried out
simulations at
\begin{equation}
\kappa\equiv J_1^\mathrm{S}/J_1^\mathrm{L}
=0, 0.5, 1.0, 1.3, 1.6, 2.0, 2.5, \infty.
\label{eq:kappalist}
\end  {equation}
When $\kappa$ is small, the lattice system dominates the whole system.
The structural phase transition is considered to occur at a high temperature.
If the room-temperature KNiCl$_3$ structure (lattice Ferri) is realized,
the spin order favors the PD state because one interaction bond is stronger
than the other two bonds in a unit triangle 
as shown in Fig. \ref{fig:frustrate}(b).
On the other hand, if the lattice structure of the $P\bar{3}c1$ symmetry
(lattice PD) is realized, the ferrimagnetic order is favored as shown in
Fig. \ref{fig:frustrate}(c).
In either case, the ground state is the ferrimagnetic state
because of finite ferromagnetic next-nearest-neighbor interactions.

\section{Method}
\label{sec:method}

We observe the following two physical quantities.
One is the sublattice order parameter:
\begin{eqnarray}
m_{\eta}^\mathrm{L}&=&\frac{1}{N_\mathrm{sub}}\sum_i\sum_{j\in \eta} 
\sigma_{ij}
\\
m_{\eta}^\mathrm{S}&=&\frac{1}{N_\mathrm{sub}}\sum_i\sum_{j\in \eta} 
S_{ij},
\end  {eqnarray}
where $\eta= \alpha, \beta, \gamma$ denotes one of three sublattices in the
triangular lattice,
and $N_\mathrm{sub}$ denotes a number of sites in one sublattice.
$m_{\eta}^\mathrm{L}$ is sublattice polarization and 
$m_{\eta}^\mathrm{S}$ is sublattice magnetization.
The other physical quantity is the $1/3$ structure factor defined using 
sublattice polarization/magnetization:
\begin{eqnarray}
f_{1/3}^\mathrm{L}&=&\frac{1}{2}\left\langle
\sum_{\eta}
(m_{\eta}^\mathrm{L})^2-
(m_{\eta}^\mathrm{L}m_{\eta'}^\mathrm{L}
+m_{\eta}^\mathrm{L}m_{\eta''}^\mathrm{L})/2
\right\rangle^{1/2}
\\
f_{1/3}^\mathrm{S}&=&\frac{1}{2}\left\langle
\sum_{\eta}
(m_{\eta}^\mathrm{S})^2-
(m_{\eta}^\mathrm{S}m_{\eta'}^\mathrm{S}
+m_{\eta}^\mathrm{S}m_{\eta''}^\mathrm{S})/2
\right\rangle^{1/2}
\\
&&(\eta\ne \eta' \ne \eta'').
\nonumber
\end  {eqnarray}
These structure factors take definite values when the Ferri or the PD
state is realized and vanish when the symmetric/paramagnetic state 
is realized at high temperatures.
$f^\mathrm{L}_{1/3}$ detects a phase boundary between an 
symmetric lattice phase and an deformed lattice phase.
$f^\mathrm{S}_{1/3}$ detects a phase boundary between a
magnetic order phase and the paramagnetic phase.

The sublattice order parameters distinguish the Ferri state and the PD state
in an deformed/ordered phase.
If one of three sublattice order parameters vanishes and the other two remain
finite, the PD state is considered to be realized.
If three sublattice order parameters take finite values, 
the Ferri state is considered to be realized.

We adopt the nonequilibrium relaxation method 
\cite{ner1,ner2,ner3}
to detect phase boundaries.
This method utilizes a relaxation function of a physical quantity in Monte
Carlo simulations.
First, an ordered state is prepared as an initial state of simulations.
We perform a simulation and obtain a relaxation function of a physical quantity.
Another simulation is performed changing the random number sequence
and another relaxation function is obtained.  We take an average of relaxation
functions over these independent Monte Carlo simulations.
This sample average is free from a systematic error due to 
averaging correlated data.
If the relaxation function of an order parameter exhibits an exponential decay,
the system is judged to be in the paramagnetic phase.
If it exhibits a converging behavior to a finite value, the system is
in the ordered phase.
The relaxation function exhibits an algebraic (power-law) decay
at the critical point.
The method has been verified to be particularly efficient in slow-dynamic 
systems with frustration and randomness.\cite{totaner1,totaner2,totaner3}

\begin{figure}[t]
\begin{center}
\includegraphics[width=8.0cm]{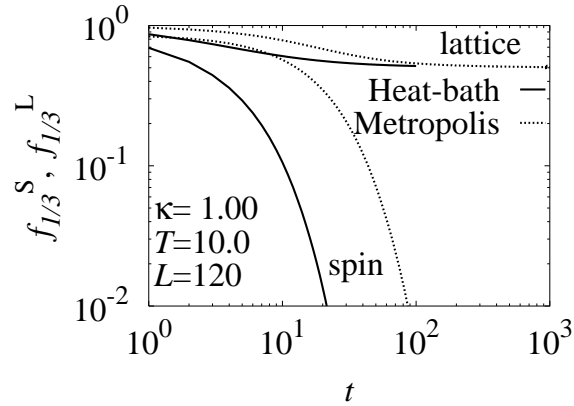}
\end{center}
\caption{A comparison between the Metropolis updating and 
the heat-bath updating.
The NER functions of the structure factors behave in the same manner while
the correlation time is about four times smaller in the heat-bath updating.
An initial state is (lattice, spin)=(Ferri, PD).
}
\label{fig:metrovsheat}
\end{figure}

In the present simulation
both spin and lattice variables at each lattice point are updated simultaneously
using the heat-bath probability.
We have checked that the Metropolis updating algorithm and the heat-bath 
updating algorithm yield the same relaxation behavior but with different
correlation time.
A simulation of the heat-bath is about four times faster as shown in
Fig. \ref{fig:metrovsheat}.

\begin{figure}[t]
\begin{center}
\includegraphics[width=8.0cm]{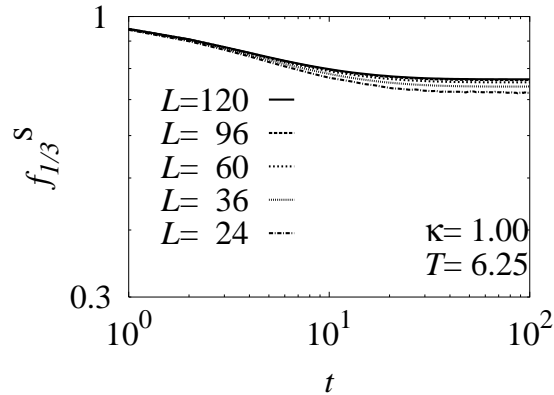}
\end{center}
\caption{
Size dependences of the structure factor of the spin.
Data of sizes smaller than $L=96$ notably deviate from data of $L=120$.
An initial state is (lattice, spin)=(Ferri, Ferri).
}
\label{fig:sizedep}
\end{figure}

The most important point in the NER method is to exclude out the finite-size 
effect from the raw relaxation function.
The method is based upon taking the infinite-size limit first.
If a relaxation function is affected by a finite-size effect,
the relaxation behavior misleads us into thinking that the temperature is 
in the paramagnetic phase even though it is in the ordered phase.
A lattice size of the present system is $L\times (L+1)\times 10L$ 
with $L$  ranging from $L=24$ to 120.
We have checked the size dependence as shown in Fig. \ref{fig:sizedep}.
The figure exhibits that the size effect appears quite early 
in the Monte Carlo step.
A system with $L=120$ is confirmed to be free from the finite-size effect 
until one hundred steps in this figure. 
For every parameter point of $\kappa$ and temperature
we have checked the size effect by comparing data of $L=120$ and
$L=96$ and confirmed a time range that the data are free from the size effect.
Results of $L=120$ are presented in this paper.

Initial conditions of the present simulations are mostly the ground-state
configuration:
the lattice structure is $P6_3cm$ and the magnetic structure is ferrimagnetic.
This combination is referred to as (lattice, spin)=(Ferri, Ferri).
We have also carried out simulations with initial configurations of
(lattice, spin)=(Ferri, PD), (Ferri, Para), (PD, PD) and (Para, Para) in order
to exclude out a possibility of initial state dependence in the final
conclusion. 

Computations are carried out by a PC cluster with 14 nodes consisting of
Pentium-4 and Athlon-XP CPU.
Total computation time is eighty days.

\section{Results}
\label{sec:results}

\subsection{The $1/3$ structure factor}

The first issue is to make it clear whether the structural phase transition
and the magnetic phase transition occur at the same temperature or not.
The NER of the $1/3$ structure factor identifies the temperature that
a lattice deformation or a magnetic order appears.

\begin{figure}[t]
\begin{center}
\includegraphics[width=8.0cm]{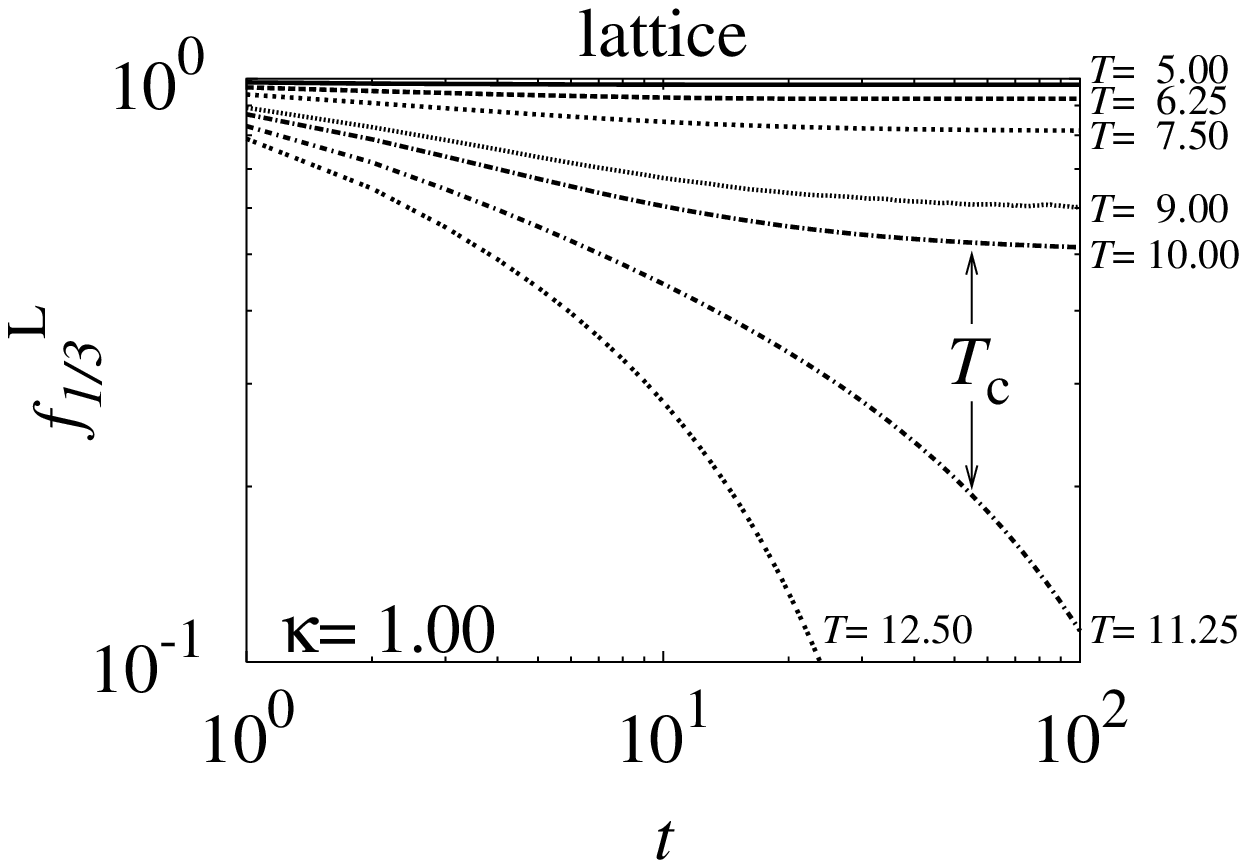}
\includegraphics[width=8.0cm]{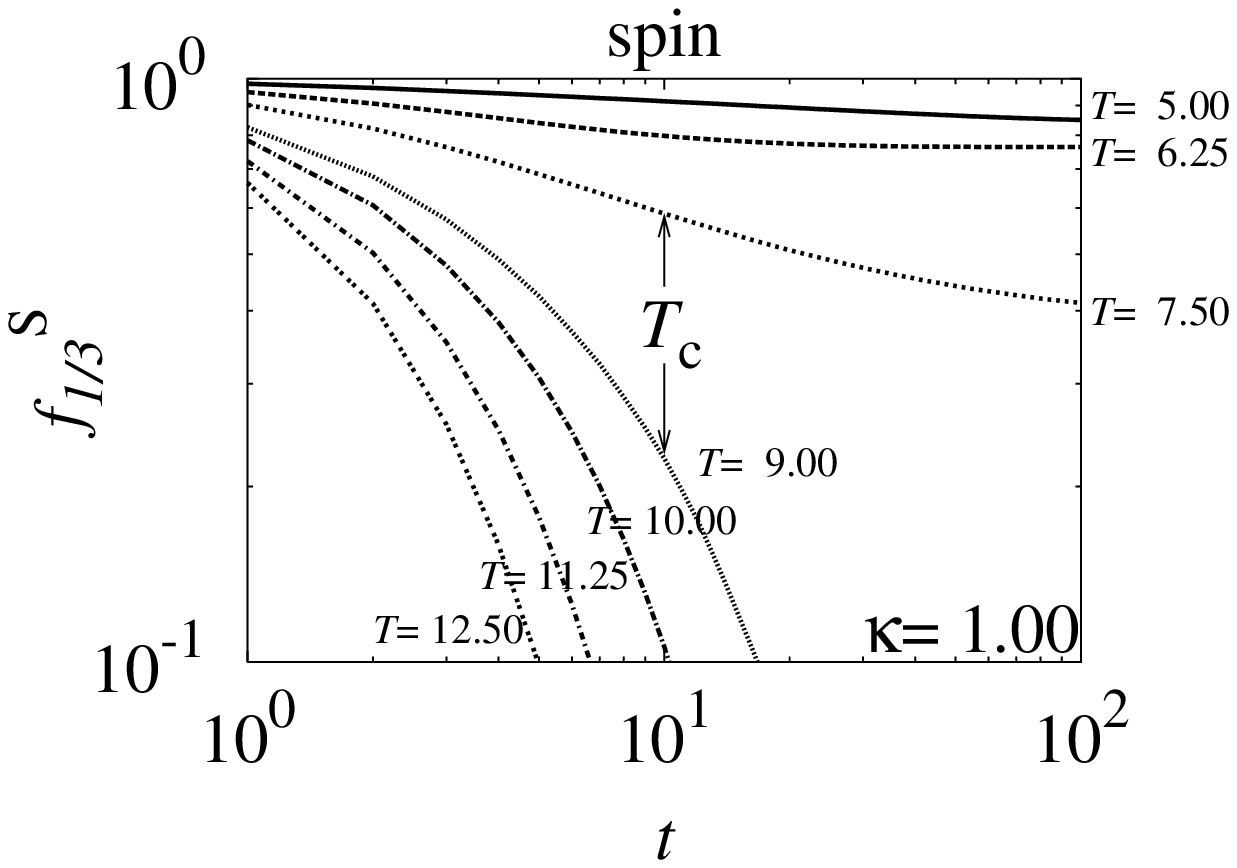}
\end{center}
\caption{NER plots of the $1/3$ structure factors of the lattice
and the spin at $\kappa=1.0$.
The temperature is as denoted aside each plot. 
The structural transition and the magnetic transition occur at different
temperatures.
}
\label{fig:f1.0}
\end{figure}

Figure \ref{fig:f1.0} shows NER plots of the structure factors
at $\kappa=1.0$.
We have started the simulation with an initial configuration of 
(lattice, spin)=(Ferri, Ferri).
This combination is realized in the ground state.
The NER of the structure factor of the lattice exhibits an
exponential decay for $T\ge 11.25$.
It exhibits a converging behavior for $T \le 10.0$.
The structural phase transition is considered to occur at 
$T_\mathrm{c}=10.625\pm 0.625$.
The spin is still paramagnetic at this temperature.
The NER of the structure factor of the spin shows that
the magnetic phase transition occurs between $T=7.5$ and $T=9.0$ 
($T_\mathrm{c}=8.25\pm 0.75$).
Therefore, two transitions occur at different temperatures.
Since the structural transition temperature is higher than that of the magnetic
transition, the lattice system is considered to dominate the
whole system at $\kappa=1.0$.

\begin{figure}[t]
\begin{center}
\includegraphics[width=8.0cm]{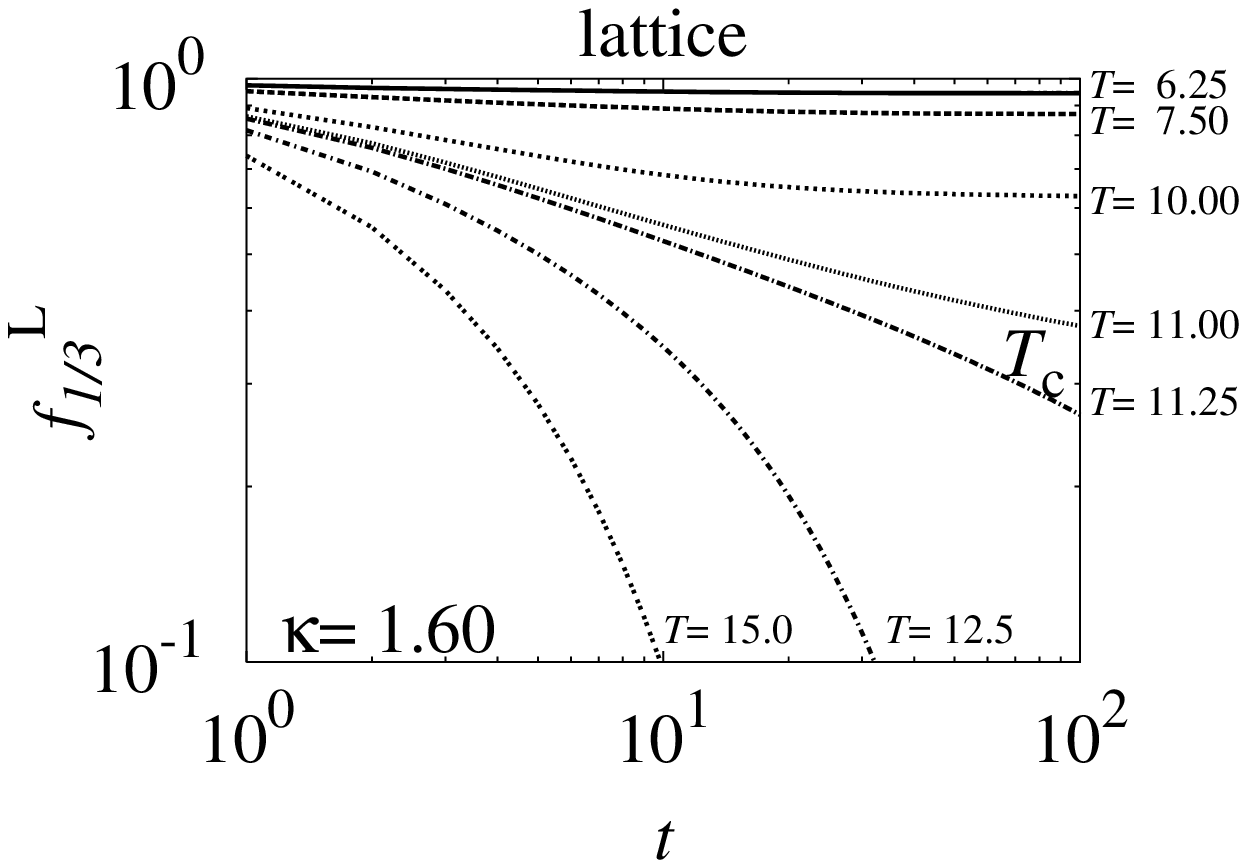}
\includegraphics[width=8.0cm]{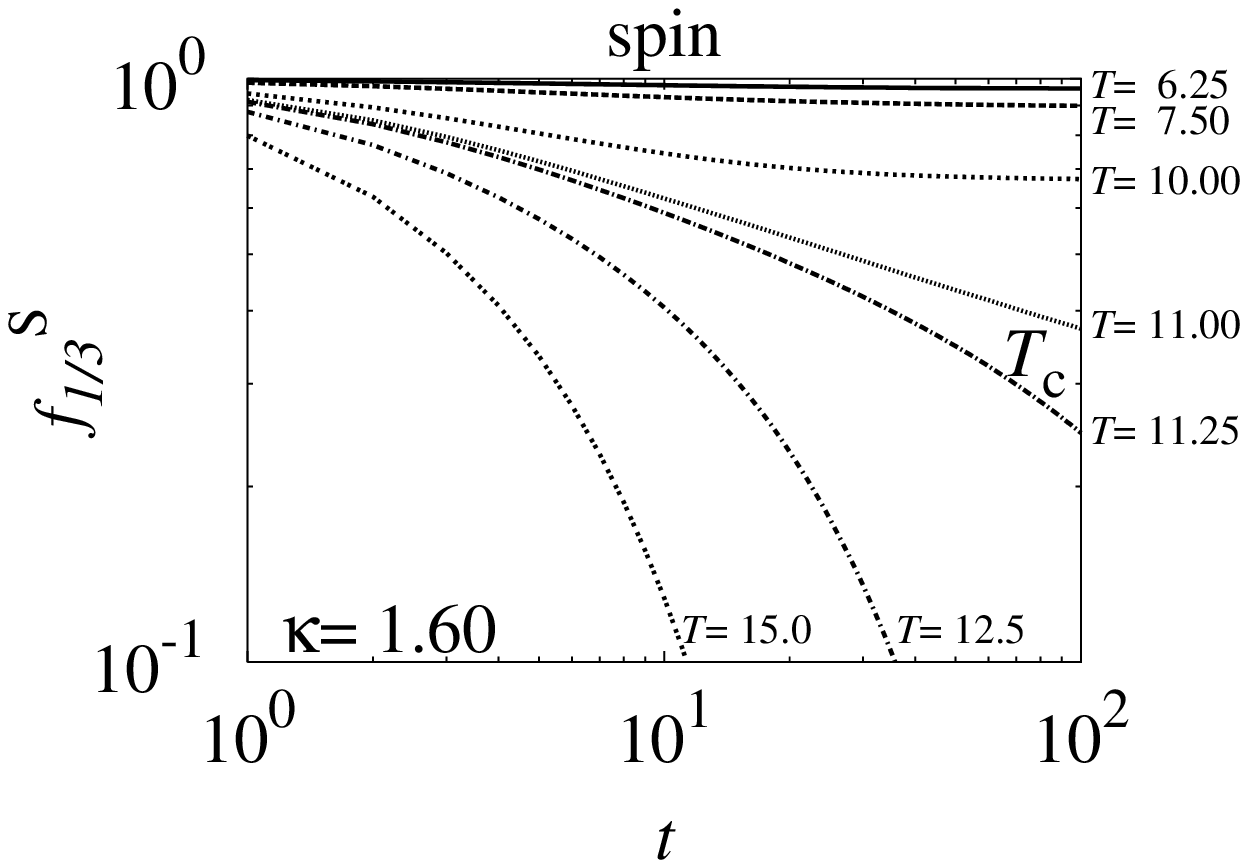}
\end{center}
\caption{
NER plots of the $1/3$ structure factors of the lattice and the spin
at $\kappa=1.6$.  The simultaneous transition occurs near $T=11.125$.
}
\label{fig:f1.6}
\end{figure}

We have carried out the same analysis changing a value of $\kappa$ as listed in
eq. (\ref{eq:kappalist}) and found that the simultaneous transition may occur
at $\kappa\simeq 1.6$ as shown in Fig. \ref{fig:f1.6}.
Both structure factors decay exponentially for $T\ge 11.25$ and 
converge to finite values for $T \le 11.00$.
The transition temperature is estimated as $T_\mathrm{c}=11.125\pm 0.125$.
For $\kappa=2.0$ and 2.5 it is found that the magnetic transition occurs
at a higher temperature than a structural transition temperature.
(Figures are not shown in this paper. 
Detailed data have been reported in the master thesis of one of the
authors (T. S.). \cite{shiramaster})
Therefore, there must be a point near $\kappa=1.6$ and $T=11.125$
that the simultaneous transition occurs.

\subsection{Sublattice polarization/magnetization}

We have observed the NER of sublattice polarization and sublattice 
magnetization in order to identify a lattice structure and a magnetic structure
at low temperatures.
It should be noticed that a sublattice may change its role of taking
magnetization of $\uparrow$, $\downarrow$ or 0 depending on samples of
independent Monte Carlo simulations.
For example, when we start a simulation from the paramagnetic state at a 
temperature that the PD state is realized,
one sublattice may take any three states of $\uparrow$, $\downarrow$ and 0
in each Monte Carlo simulation.
Therefore, a simple average of sublattice magnetization over the independent 
Monte Carlo simulations yields an average
of magnetization over sublattices: magnetization is obtained.
In order to distinguish the above ambiguity, we collect data of sublattice
magnetization depending on their value.
Sublattice magnetization whose value is the smallest among three is 
stored in $m_\mathrm{min}^\mathrm{S}$.
For example, a spin-down sublattice ($\downarrow$) in the PD state
contributes to this variable.
That of the largest value is stored in $m_\mathrm{max}^\mathrm{S}$ 
($\uparrow$ in PD),
and the rest is stored in $m_\mathrm{mid}^\mathrm{S}$ (0 in PD).

\begin{figure}[t]
\begin{center}
\includegraphics[width=8.0cm]{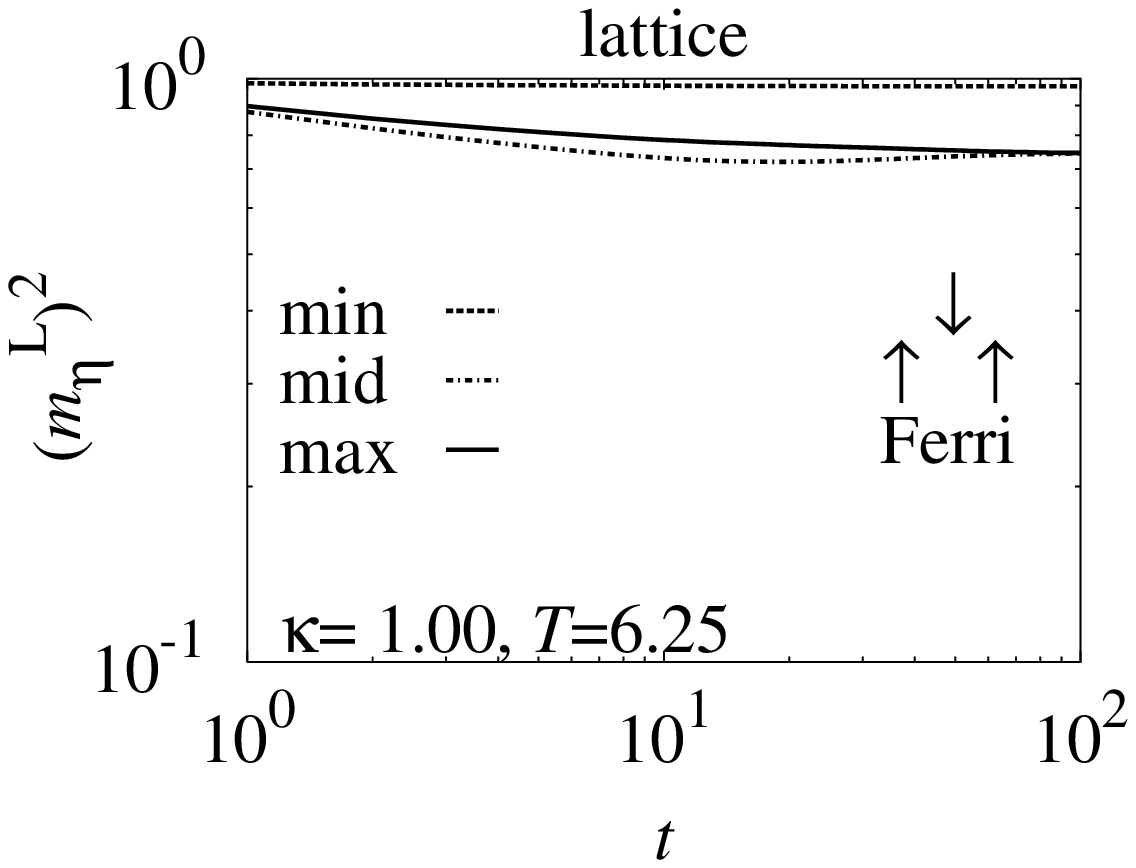}
\includegraphics[width=8.0cm]{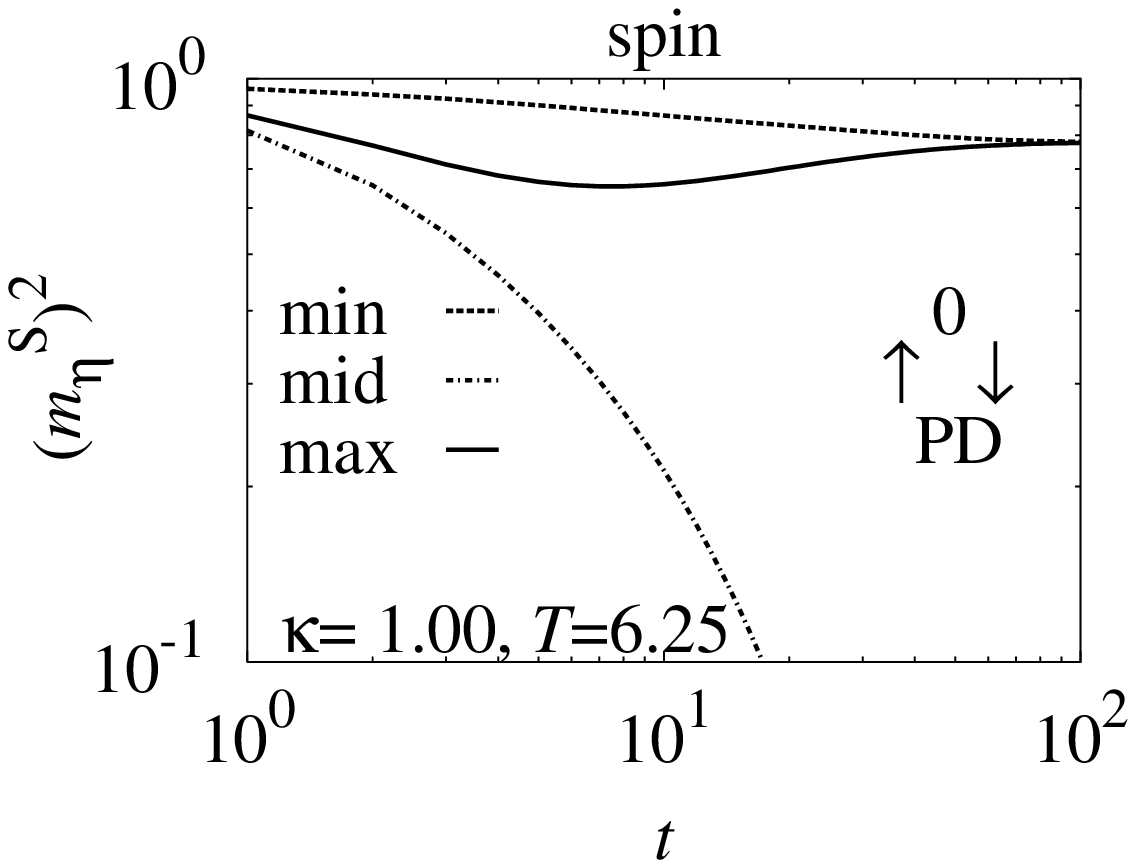}
\end{center}
\caption{
NER plots of squared sublattice polarization (lattice) and squared sublattice
magnetization (spin) at $\kappa=1.0$ and $T=6.25$.
The lattice Ferri state and the magnetic PD state are realized.
}
\label{fig:m1.0}
\end{figure}
Figure \ref{fig:m1.0} shows the NER of squared sublattice 
polarization/magnetization at $\kappa=1.0$ and $T=6.25$.
The initial state is (lattice, spin)=(Ferri, Ferri).
Three sublattice polarization converge to finite values.
Here, $m_\mathrm{min}^\mathrm{L}$ corresponds to $\downarrow$-state and its
squared value is largest among three.
$m_\mathrm{mid}^\mathrm{L}$ and $m_\mathrm{max}^\mathrm{L}$ correspond to 
$\uparrow$-state and converge to the same value.
A realized state is considered to be the lattice Ferri ($P6_3cm$).
The NER of squared sublattice magnetization suggests that the magnetic
structure is the PD state.
$m_\mathrm{min}^\mathrm{S}$ and $m_\mathrm{max}^\mathrm{S}$ respectively
correspond to the
$\downarrow$-state and the $\uparrow$-state and converge to a finite value.
$m_\mathrm{mid}^\mathrm{S}$ vanishes exponentially.
Therefore, it is concluded that (lattice, spin)=(Ferri, PD) state is realized
at $\kappa=1.0$ and $T=6.25$.
We have verified this result by starting simulations with (Ferri, PD) state,
(PD, PD) state and (Ferri, Para) state.
Every relaxation plot has drawn out the same conclusion.\cite{shiramaster}

\begin{figure}[t]
\begin{center}
\includegraphics[width=7.3cm]{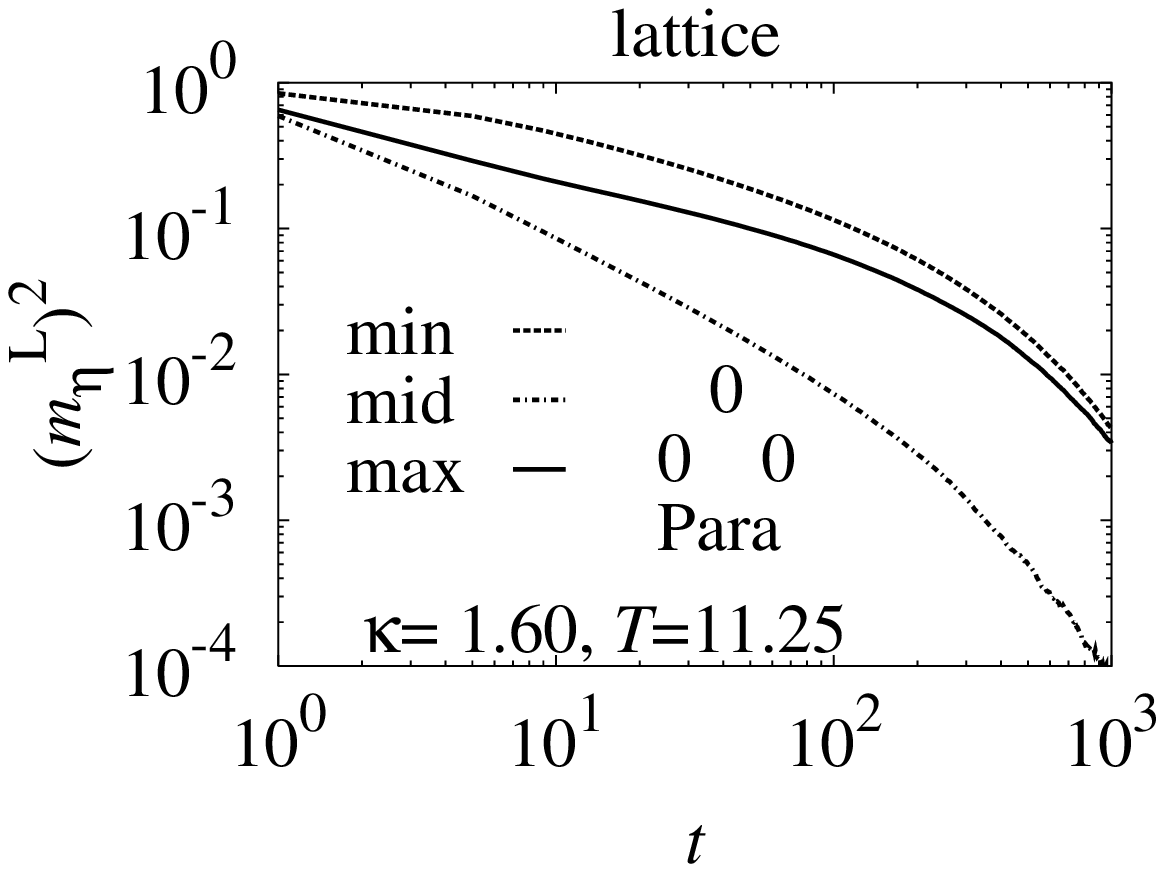}
\includegraphics[width=7.3cm]{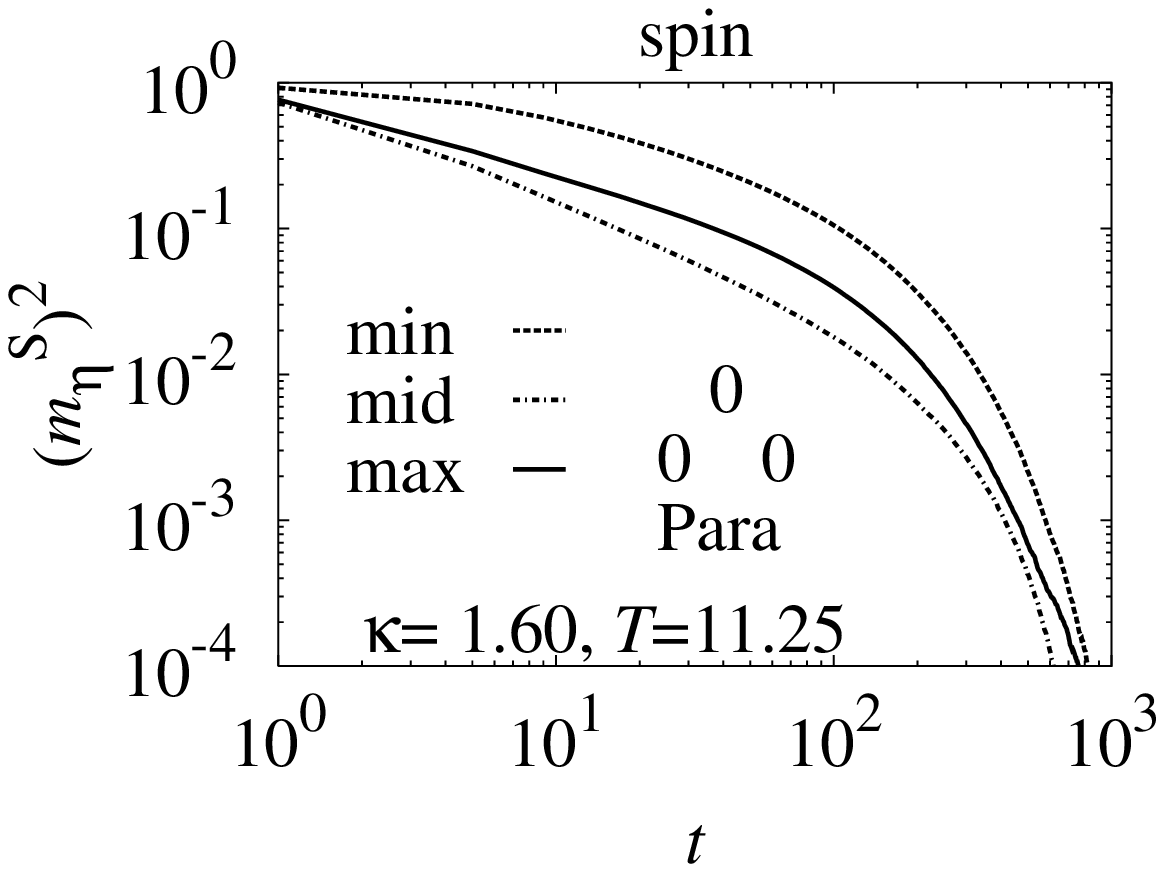}
\includegraphics[width=7.3cm]{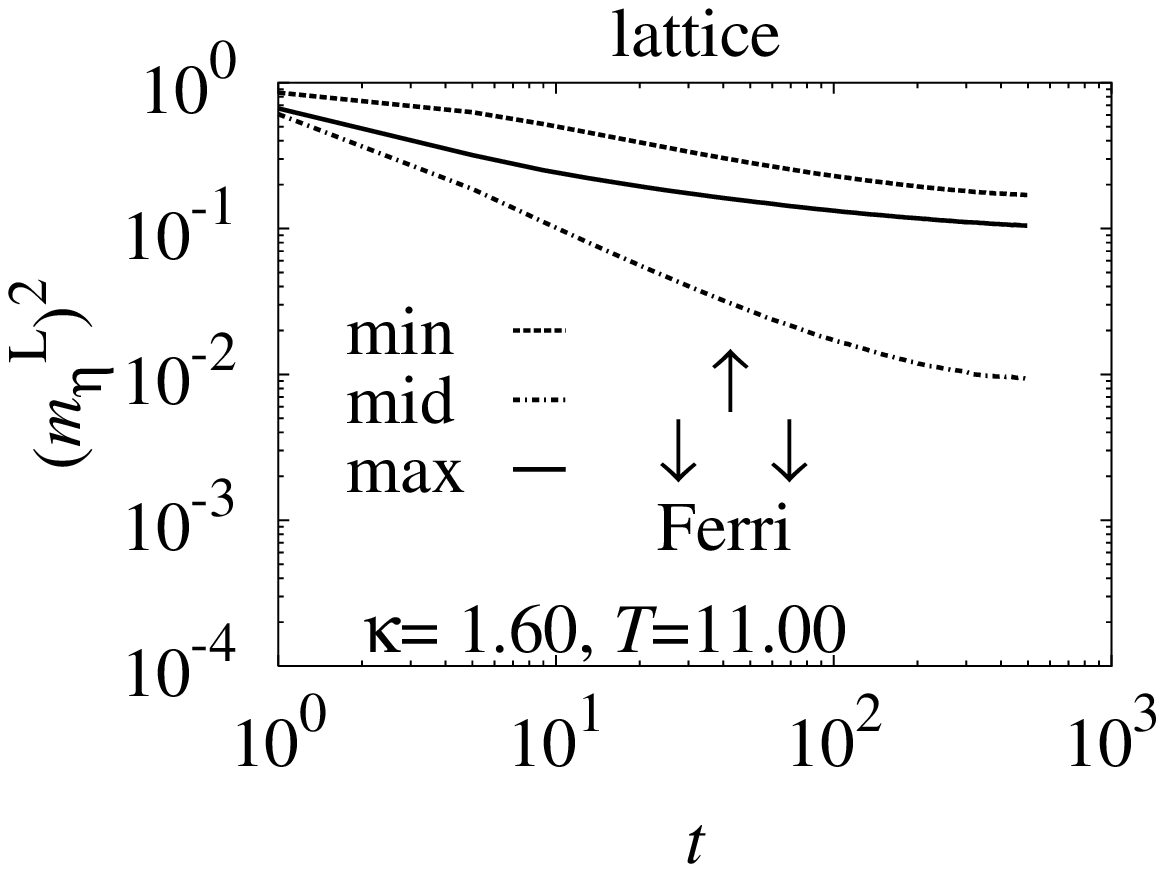}
\includegraphics[width=7.3cm]{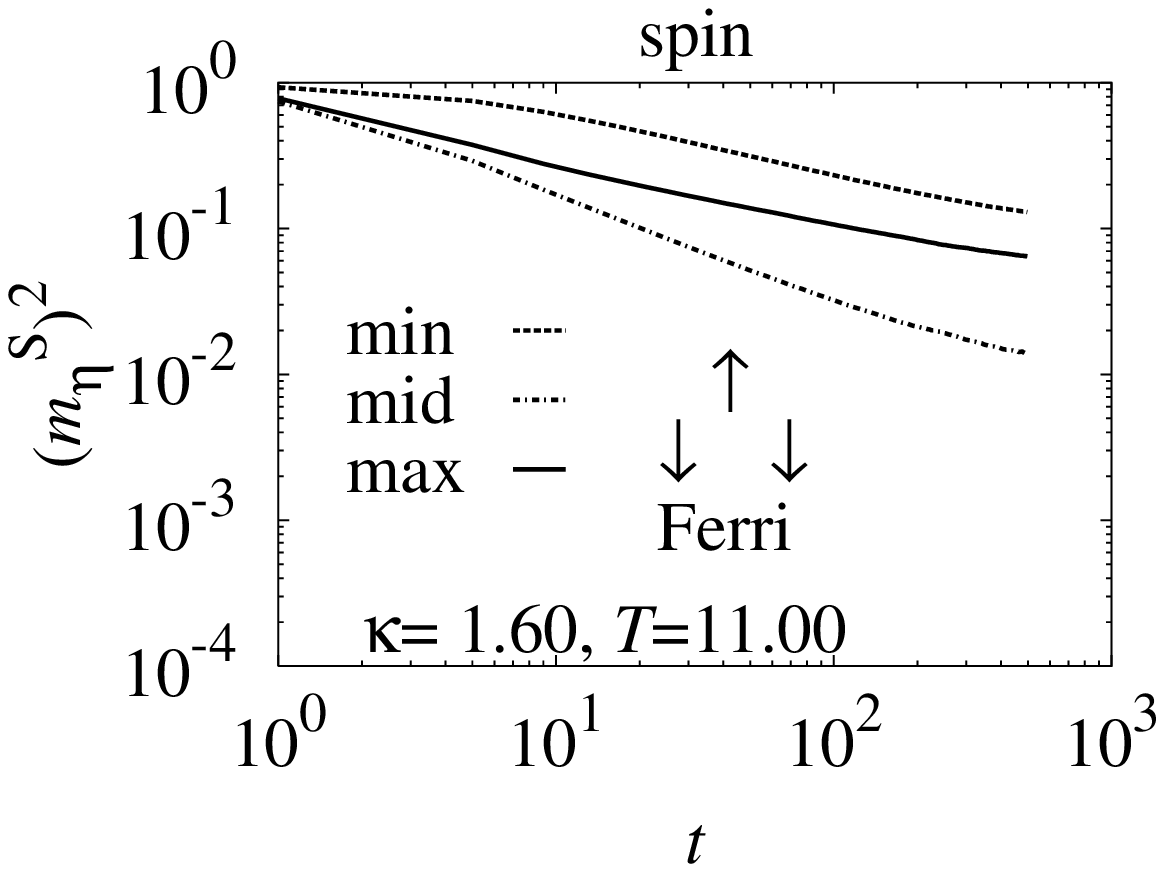}
\end{center}
\caption{The NER of sublattice polarization/magnetization at
$\kappa=1.6$. The realized state is (lattice, spin)=(Para, Para) at $T=11.25$,
while it is (lattice, spin)=(Ferri, Ferri) at $T=11.00$.}
\label{fig:k1.6}
\end{figure}

Figure \ref{fig:k1.6} shows the NER of polarization/magnetization at
$\kappa=1.6$ just above and below the simultaneous transition temperature.
All the relaxation functions decay exponentially at $T=11.25$.
This evidence suggests that both lattice and spin take symmetric/paramagnetic
configurations at this temperature.
Contrary, the relaxation functions converge to finite values at $T=11.00$.
The (lattice, spin)=(Ferri, Ferri) state is considered to be realized.
This is the ground state configuration.
A direct transition from the high temperature phase to the ground-state phase
has occurred.
Intermediate phases vanish at this point.

\section{Discussion}
\label{sec:discussion}

\begin{figure}[t]
\begin{center}
\includegraphics[width=8.0cm]{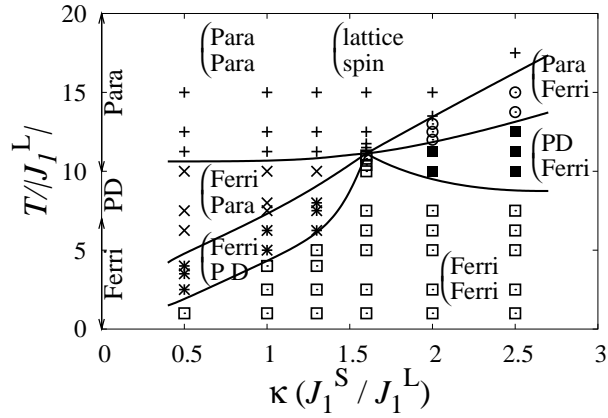}
\end{center}
\caption{A phase diagram of the present model. 
Symbols discriminate phases as denoted in brackets.
}
\label{fig:phase}
\end{figure}

Putting all the NER results together we obtain a $T$-$\kappa$ 
phase diagram of the present model as shown in Fig. \ref{fig:phase}.
When there is no spin degree of freedom at $\kappa=0$, 
successive structural phase transitions occur 
from the symmetric phase ($P6_c/mmc$) to the lattice PD phase ($P\bar{3}c1$)
and then to the lattice Ferri phase ($P6_3cm$).
Contrary, when there is no lattice degree of freedom at $\kappa=\infty$,
we have also confirmed that successive magnetic transitions occur from 
the paramagnetic phase to the PD phase and then to the ferrimagnetic phase.

In case where both lattice and spin exist, a structural
transition and a magnetic transition occur.  
If an energy scale of the lattice is larger than 
that of the spin ($\kappa \le 1.3$), the structural transition occurs at a 
higher temperature than magnetic transition temperatures.
At this transition temperature a lattice changes its structure directly from
the symmetric one to the ground-state one.
The intermediate phase does not appear.
The successive magnetic transitions occur at lower temperatures.
This interesting phenomenon can be explained by the following scenario.
A key concept is relaxation of frustration.

A system with a larger energy scale becomes a master and 
the other one becomes a slave.
When the structural transition occurs at a higher temperature, 
the lattice is a master and the spin is a slave.
The spin as a slave serves a master relaxing frustration of the lattice 
by taking the PD configuration.
As explained in Fig. \ref{fig:frustrate}, the spin favors the PD configuration
when the lattice takes the Ferri structure, and vice versa.
Since the temperature is still high, the PD spin configuration cannot be
a long-range order.
However, the short-range order is sufficient to relax frustration.
The lattice can take the ground-state configuration without experiencing the 
intermediate PD phase.

A phase boundary of the structural transition does not depend on $\kappa$ 
when $\kappa \le 1.3$, and smoothly interpolate to the phase boundary
between Para and PD at $\kappa=0$.
It is speculated that the structural PD phase suddenly vanishes by an
introduction of an infinitesimal spin degree of freedom.
When the temperature is lowered, the successive magnetic transitions occur.
There is no slave relaxing frustration of the spin variable.
Therefore, the intermediate phase appears.

Roles of the lattice and the spin are completely exchanged when an energy
scale of the spin part exceeds that of the lattice part for $\kappa > 1.6$.
The spin becomes a master and the lattice becomes a slave.
The magnetic transition occurs at a higher temperature and the ground-state
configuration (ferrimagnetic state) is realized by an assistance of the lattice
system.
The successive structural transitions occur at lower temperatures.

The simultaneous magneto-dielectric transition occurs when two energy scales
coincide.
Either system relaxes frustration of the other system cooperatively.
Both realize the ground-state configuration at the same temperature.
Intermediate phases of both systems vanish.
The simultaneous transition temperature corresponds to the real temperature
as 
\begin{equation}
(T_\mathrm{c}/|J_1^\mathrm{L}|)\times J_1^\mathrm{S} / \kappa=
11.125\times 2.5 / 1.6 \simeq 17\mathrm{K}.
\end  {equation}
It is almost half the experimental result of 37K.
This underestimate is due to adopting the large reductions of
exchange integrals: $\Delta J_{(c,1,2)}^\mathrm{S}=0.2J_{(c,1,2)}^\mathrm{S}$.
The exchange integrals can be reduced to 60\% of the original value.
We made the reduction large in order to observe the spin-lattice cooperative
phenomenon easily.
However, 
it decreases an energy scale of the spin part, and consequently decreases
the magnetic transition temperature.
There is also an ambiguity in an experimental estimate of $J_1^\mathrm{S}$.
Therefore, we consider that our estimate of the transition temperature
is not inconsistent with the experimental result.

This scenario possibly explains the simultaneous transition in RbCoBr$_3$.
It is also speculated that there is no further magnetic transition at 
lower temperatures.
The ground-state magnetic state (possibly the ferrimagnetic state) is 
considered to be realized below the simultaneous transition temperature.
Recently, it is reported that a single-crystal neutron diffraction experiment
has revealed that all three sublattice magnetization are ordered
at 37K.\cite{nishiwaki}
This experimental evidence supports our speculation of an appearance of the
ferrimagnetic state at the simultaneous transition point.

In this paper we have introduced a model Hamiltonian which explains 
magneto-dielectric phase transitions in ABX$_3$-type compounds.
It is supposed that any deformation of a lattice decreases the magnetic
exchange interaction.
The model is made as simple as possible just to understand the phenomenon
qualitatively.
Choices of the system parameters are not realistic.
Therefore, there remains much to improve in order to compare theoretical
results with experimental results quantitatively.

An essential point of this phenomenon is relaxation of frustration 
by introducing two systems: lattice and spin.
Frustration finds a way to be relaxed by utilizing the other system.
If it is relaxed, the intermediate phase like the PD phase vanishes.
It can be considered that the intermediate phase is a product of frustration
that has nowhere to go.

\acknowledgment

The authors would like to thank Professor Tetsuya Kato for guiding their 
interests to the present phenomenon and for fruitful discussions.
The author T.N. also thank Yoichi Nishiwaki for fruitful discussions.
The use of random number generator RNDTIK programmed by
Professor Nobuyasu Ito and Professor Yasumasa Kanada is gratefully acknowledged.

\end{document}